\documentclass[aps,pra,reprint,groupedaddress,showpacs,showkeys,superscriptaddress]{revtex4-1}

\usepackage[pagebackref=false,colorlinks,linkcolor=red,citecolor=magenta]{hyperref}
\usepackage{graphicx}
\begin{document}


\title{
An analogue model for controllable Casimir radiation in a nonlinear cavity with amplitude-modulated pumping:
Generation and quantum statistical properties 
}
\author{Ali Motazedifard} 
\email{motazedifard.ali@gmail.com}
\address{Department of Physics, Faculty of Science, University of Isfahan, Hezar Jerib, 81746-73441, Isfahan, Iran}
\author{M. H. Naderi} 
\email{mhnaderi@phys.ui.ac.ir}
\author{R. Roknizadeh}
\email{rokni@sci.ui.ac.ir}
\address{Department of Physics, Faculty of Science, University of Isfahan, Hezar Jerib, 81746-73441, Isfahan, Iran}
\address{Quantum Optics Group, Department of Physics, Faculty of Science, University of Isfahan, Hezar Jerib, 81746-73441, Isfahan, Iran}
\date{\today}
\begin{abstract}
We present and investigate an analogue model for a controllable photon generation via the dynamical Casimir effect (DCE) in a cavity containing a degenerate optical amplifier (OPA) which is pumped by an amplitude-modulated field. The time modulation of the pump field in the model OPA system is equivalent to a periodic modulation of the cavity length, which is responsible for the generation of the Casimir radiation. By taking into account the rapidly oscillating terms of the modulation frequency, the effects of the corresponding counter-rotating terms (CRTs) on the analogue Casimir radiation emerge clearly. We find that the mean number of generated photons and their quantum statistical properties exhibit oscillatory behaviours, which are controllable through the modulation frequency as an external control parameter. We also recognize a new phenomenon, the so-called  ``Anti-DCE ", in which pair photons can be coherently annihilated due to the time-modulated pumping. We show that the Casimir radiation exhibits quadrature squeezing, photon bunching and super-Poissonian statistics which are controllable by modulation frequency. We also calculate the power spectrum of the intracavity light field. We find that the appearance of the side bands in the spectrum is due to the presence of the CRTs.
\end{abstract}
\pacs{42.50.Ct, 42.50.Pq, 42.50.Ar, 42.65.Yj}
\keywords{Dynamical Casimir Effect, Analogue Models, Nonlinear Cavity, Optical Parametric Amplifier, Nonclassical Properties }

\maketitle

\section{Introduction}

One of the most important and interesting predictions of quantum field theory is that the quantum vacuum state is not truly empty, but instead filled with virtual particles which are continuously created and annihilated due to the zero-point fluctuations \cite{1}. In quantum electrodynamics these fluctuations lead to a number of measurable effects in the realm of microscopic physics, such as the van der Waals force between atoms in vacuum, natural widths of spectral lines, Lamb shift, and anomalous magnetic moment of electron \cite{1}. In macroscopic level, the vacuum fluctuations manifest also themselves in a variety of observable mechanical phenomena and the archetype of them is the so-called static Casimir effect \cite{2,3}. This effect which arises from the pressure that virtual photons exert on stationary boundaries, has been observed in different experiments \cite{4,5} and plays a very important role both in fundamental physics investigations as well as in understanding the basic limits of nanomechanical technologies \cite{6,7}. 

Another manifestation of the zero-point fluctuations is the creation of real particles out of the vacuum induced by the interaction with dynamical external constraints \cite{8}. In this effect, energy is transferred from the external field to the zero-point fluctuations transforming them into real particles. Although a material boundary is somewhat analogous to an external constraint, there is no particle creation from the vacuum in the case of static boundaries (static external constraints). However, if the boundaries are nonstationary, and the boundary conditions depend on time, there is particle creation from the vacuum in addition to the Casimir force. Such quantum vacuum amplification effect is commonly referred to as the dynamical Casimir effect (DCE) \cite{9,10}(for an extensive recent review, see \cite{11}). Whereas the static Casimir effect arises from a mismatch of vacuum modes in space, the DCE originates from a mismatch of vacuum modes in time domain. Examples of the DCE range from cosmology, such as particle creation in a time-dependent cosmological background \cite{12,13,14} to cavity QED, such as photon generation in Fabry-Perot cavities with moving mirrors \cite{15,16,17,18,19,20,21}. There is an extensive literature on various aspects of the creation of photons due to the DCE using various theoretical methods. For example, photon creation was described based on a Hamiltonian approach \cite{22}. Cavities with the insertion of dispersive mirrors or a slab with a time-dependent dielectric permittivity were considered \cite{23,24,Microscopic toy model,Prospects for observing}. Multiple-scale analysis was applied to the calculation of the flux of created particles \cite{18}. Photon creation in a harmonically oscillating one-dimensional cavity with mixed boundary conditions has been analyzed \cite{25}. The DCE has also been investigated in the context of cavity QED \cite{26,27}, in Bose-Einstein condensates \cite{28}, in excition-polariton condensates \cite{29}, in superconducting circuits\cite{30}, and in a quantum-well assisted optomechanical cavity \cite{31}. Besides its intrinsic importance as a direct proof of the vacuum fluctuations, various practical applications of the DCE has been proposed, e.g., for generation of photons with nonclassical properties \cite{32,33,34,35}, for high precision optical interferometry\cite{36}, for generation of atomic squeezed states \cite{37}, for multipartite entanglement generation in cavity networks\cite{38}, and for quantum communication protocols \cite{39}.

From the experimental point of view, the detection of the DCE induced by moving boundaries is a serious problem. The main difficulty is largely due to the challenging prerequisite of generating appreciable mechanical frequencies of moving boundaries (of the order of GHz) to obtain a detectable number of photons. Although there are some experimental proposals for the detection of the DCE involving real mechanical motion of boundaries (see for instance \cite{40}), several analogue systems have been proposed for observing the Casimir radiation based on simulating moving boundaries by considering material bodies with time-dependent electromagnetic properties. The first analogue system consisting of a nonlinear optical medium with time-dependent refractive index was introduced in \cite{9} and later on, its experimental implementation was reported \cite{42}. In \cite{43} the Casimir radiation within a cavity containing a thin semiconducting film with time-dependent conductivity and centered at the middle of the cavity has been investigated. An explicit analogy between photon emission of a coherently pumped thin nonlinear crystal of type ${\chi ^{(2)}}$ inside a cavity and the photon generation via the DCE have been established \cite{44}. In \cite{45} an experimental layout for measuring the DCE has been proposed where the optical length of a cavity is varied in time by means of a train of ultrashort pulses that modulate the effective refractive index of the cavity material. By using a varying boundary condition in a superconducting waveguide \cite{46,47} which simulates a single moving mirror, the first experimental demonstration of the DCE has recently been reported \cite{48}. Moreover, a second observation of the DCE has been reported \cite{49} based on periodical changes in the index of refraction of a microwave cavity with a Josephson metamaterials. 

Motivated by the above-mentioned investigations on analogue models for the DCE, in the present contribution we propose a model system for the generation of the Casimir radiation in a cavity containing a degenerate optical parametric amplifier (OPA) which is pumped by an amplitude-modulated field according to the harmonic law
\begin{equation}
{\varepsilon _p}(t) = \varepsilon  + \eta \cos \Omega t,
\label{pump}
\end{equation}
where $\varepsilon$ is a constant amplitude and $\eta$, $\Omega$ are, respectively, the amplitude and frequency of modulation. The temporal change in pump amplitude leads to a modulation of the cavity length which is equivalent to an apparent displacement of the cavity mirrors. In all of the previous investigations carried out on the DCE, a kind of rotating wave approximation (RWA) has been applied to neglect the contribution of rapidly oscillating terms of the modulation frequency( see for instance \cite{50,51,52,53,DCE in kerr cavity}). The main feature of our treatment is that we release this limitation to include the corresponding counter rotating terms (CRTs). We will find that taking into account the CRTs in the model under consideration leads to an interference between the constant amplitude, $\varepsilon$, and the oscillatory term. The total number of generated photons consists of three contributions, i.e., constant pumping, time varying pumping, and their interference which originates from CRTs. Due to this interference, the mean number of generated photons and their quantum statistical properties exhibit oscillatory behaviours which are controllable through the modulation frequency $\Omega$. 

The paper is structured as follows. In section. \ref{sec2} we describe the model system and show the analogy between the time modulation of the pump
and the time modulation of the cavity length in the conventional DCE. We also solve analytically the quantum Langevin equations for the cavity-field operators. We calculate the mean number of generated Casimir photons and analyse its temporal behaviour in section. \ref{sec3}. We study the quadrature squeezing, photon counting statistics and the power spectrum of the generated Casimir radiation in section. \ref{sec4}. Finally, we summarize the main results of the paper in section. \ref{sec5}.

\section{\label{sec2}DESCRIPTION OF THE SYSTEM}
\begin{figure}
\begin{center}
\includegraphics[scale=1]{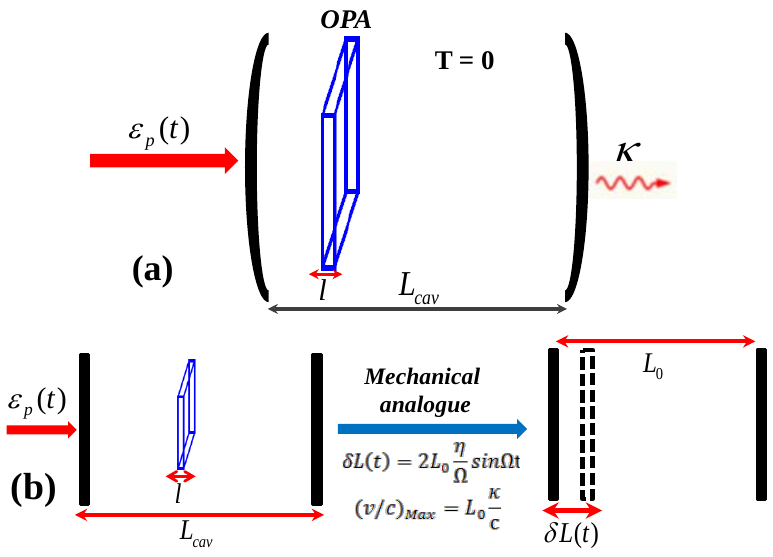}
\end{center}
\caption{(Color online) (a) Schematic representation of a Fabry-Perot cavity containing an OPA which is pumped by an amplitude-modulated field. (b) The analogy between the cavity with time modulated pumping and the cavity with oscillating mirror} 
\label{Fig1}
\end{figure}
As schematically shown in Fig.\ref{Fig1}(a), we consider a system composed of a sub-threshold degenerate OPA inside a Fabry-Perot cavity with fixed mirrors at zero temperature. We assume that the OPA is pumed by a coupling field whose amplitude is time-modulated according to Eq. (\ref{pump}). We also take into account the dissipation of the cavity field to the environment with a decay rate $\kappa$. We show that this non-stationary cavity QED results in the DCE. In a degenerate OPA a pump photon of frequency $\omega_p$ is down- converted into a pair of identical photons of frequency $\omega_0$. The Hamiltonian describing the system in the interaction picture is given by
\begin{equation}
{\hat H_I} = \frac{i}{2}{\varepsilon _p}(t)\left( {{{\hat a}^{\dag 2}} - {{\hat a}^2}} \right),
\label{Eq1}
\end{equation}
where $\hat a$(${{\hat a}^\dag }$) is the annihilation (creation) operator of the cavity field. Here, it should be emphasized that in the above Hamiltonian we will not make use of the RWA for the term $\cos \Omega t$ in the amplitude $\varepsilon_p(\ t)$ and we want to examine the influence of the corresponding CRTs on the generation and quantum statistical properties of the Casimir photons. Before proceeding to discuss the dynamics of the  system under consideration, it is worthwhile to beriefly describe the analogy between the time modulation of the pump and the time modulation of the cavity length in the conventional DCE. Following the same method as used in \cite{44}, it is easy to show that the time-modulated nonlinear parametric interaction in the model system, described by the Hamiltonian (\ref{Eq1}), is equivalent to an apparent periodic displacement of the cavity mirrors which is given by $\delta L(t) = 2{L_0}(\eta /\Omega )\sin \Omega t$, where the effective length of cavity is ${L_0} = {L_{cav}} + l({n_{OPA}} - 1)$, with $l$ and $\ n_{OPA}$ as the length and the refractive index of the nonlinear crystal, respectively (see Fig.\ref{Fig1}(b)). Furthermore, the maximum of the effective speed of the mirror is obtained as ${(v/c)_{max}} = {\left. {\delta \dot L(t)/c} \right|_{\eta  = \kappa /2}} = \kappa {L_0}/c$ (for example, if ${L_0} \sim cm$ and $\kappa  \sim10^4-10^6$ Hz, then ${(v/c)_{max}}$ is of order $10^{-4}-10^{-2}$). As we will show latter, in comparison to the analogue model introduced in \cite{44}, where the refractive index of nonlinear crystal coated on the mirror is modulated, our analogue model has the advantage of providing the possibility of controlling the generation of the Casimir photons and their quantum statistical properties by adjusting the frequency of modulation, $\Omega$. This controllability results from taking into account the CRTs in the amplitude ${\varepsilon _p}(t)$.

The quantum Langevin equations of the system, taking into account the cavity dissipation, can be written as
\begin{equation}
\begin{array}{l}
\frac{{d\hat a}}{{dt}} =  - \frac{\kappa }{2}\hat a + {\varepsilon _p}(t){{\hat a}^\dag } + {{\hat F}_c}(t),\\
\frac{{d{{\hat a}^\dag }}}{{dt}} =  - \frac{\kappa }{2}{{\hat a}^\dag } + {\varepsilon _p}(t)\hat a + \hat F_c^\dag (t),
\end{array}
\label{Eq2}
\end{equation}
where ${{\hat F}_c}(t)$ is the noise operator associated with the dissipation of the cavity field with zero mean value, i.e.,$\left\langle {{{\hat F}_c}(t)} \right\rangle  = 0$. For a cavity mode damped by a vacuum reservoir, the noise operator in zero temperature satisfies the Markovian correlation functions:
\begin{equation}
\left\langle {{{\hat F}_c}(t)\hat F_c^\dag (t')} \right\rangle  = \frac{\kappa }{2}\delta (t - t'),
\label{Eq3}
\end{equation}
\begin{equation}
\left\langle {\hat F_c^\dag (t){{\hat F}_c}(t')} \right\rangle  = \left\langle {{{\hat F}_c}(t){{\hat F}_c}(t')} \right\rangle  = \left\langle {\hat F_c^\dag (t)\hat F_c^\dag (t')} \right\rangle= 0 .
\label{Eq4}
\end{equation}
By defining the operators $\hat X: = {{\hat a}^\dag } + \hat a$, $\hat Y: = {{\hat a}^\dag } - \hat a$, and
${{\hat F}_ \pm }: = \hat F_c^\dag (t) \pm {{\hat F}_c}(t)$ the quantum Langevin equations can be written as
\begin{equation}
\begin{array}{l}
\hat {\dot X} = \left( {{\varepsilon _p}(t) - \frac{\kappa }{2}} \right)\hat X + {{\hat F}_ + }(t),\\
\hat {\dot Y} =  - \left( {{\varepsilon _p}(t) + \frac{\kappa }{2}} \right)\hat Y + \hat F_ - ^\dag (t) .
\end{array}
\label{Eq5}
\end{equation}
The solution of the set of Eqs. (\ref{Eq5}) is given by 
\begin{equation}
\begin{array}{l}
\hat X(t) = {f_1}(t)\left( {\hat X(0) + \int_0^t {dt'{{\hat F}_ + }(t'){g_1}(t')} } \right),\\
\hat Y(t) = {f_2}(t)\left( {\hat Y(0) + \int_0^t {dt'{{\hat F}_ - }(t'){g_2}(t')} } \right),
\end{array}
\label{Eq6}
\end{equation}
where
\begin{equation}
\begin{array}{l}
{f_1}(t) = {{\mathop{\rm \textit{e}}\nolimits} ^{(\varepsilon  - \kappa /2)t + \tilde \eta \sin (\Omega t)}} ,\qquad 
{g_1}(t) = {{\mathop{\rm \textit{e}}\nolimits} ^{ - (\varepsilon  - \kappa /2)t - \tilde \eta \sin (\Omega t)}} ,\\
{f_2}(t) = {{\mathop{\rm \textit{e}}\nolimits} ^{ - (\varepsilon  + \kappa /2)t - \tilde \eta \sin (\Omega t)}} ,\qquad 
{g_2}(t) = {{\mathop{\rm \textit{e}}\nolimits} ^{(\varepsilon  + \kappa /2)t + \tilde \eta \sin (\Omega t)}} .
\end{array}
\label{Eq8}
\end{equation}
with $\tilde \eta =\eta /{\Omega }$. Thus, the field operators are given by the following expressions
\begin{equation}
\begin{array}{l}
\hat a(t) = \frac{1}{2}\left( {{f_1}(t)\hat X(0) - {f_2}(t)\hat Y(0) + } \right.\\
\qquad  \left. {{f_1}(t)\int_0^t {dt'{{\hat F}_ + }(t'){g_1}(t')}  - {f_2}(t)\int_0^t {dt'{{\hat F}_ - }(t'){g_2}(t')} } \right) ,\\
{{\hat a}^\dag }(t) = \frac{1}{2}\left( {{f_1}(t)\hat X(0) + {f_2}(t)\hat Y(0) + } \right.\\
\qquad\left. {{f_1}(t)\int_0^t {dt''{{\hat F}_ + }(t''){g_1}(t'')}  + {f_2}(t)\int_0^t {dt''{{\hat F}_ - }(t''){g_2}(t'')} } \right) .
\end{array}
\label{Eq7}
\end{equation}
\section{\label{sec3}MEAN NUMBER OF GENERATED PHOTONS}
\begin{figure}
\begin{center}
\includegraphics[scale=1.02]{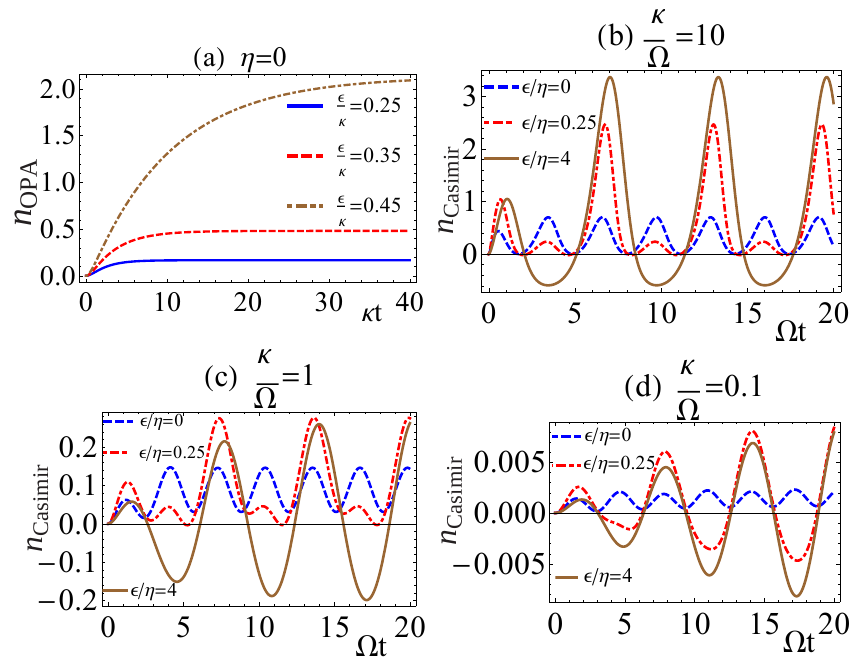}
\end{center}
\caption{(Color online) (a) Mean number of generated photons in the absence of time modulation ($\eta = 0$) versus the dimensionless time $\kappa t$ for different values of $\varepsilon/\kappa$. (b), (c) and (d) Mean number of Casimir photons versus the dimensionless time $\Omega t$, for different values of $\kappa/\Omega$ and $\varepsilon/\eta$. Here, we have set $\kappa/ 2\pi=10^4$Hz.
} 
\label{Fig2}
\end{figure}

We are now in a position to calculate the mean number of intracavity photons by using the analytic solution given by Eq.(\ref{Eq7}). Assuming that the cavity field is initially prepared in the vacuum state,$\left| 0 \right\rangle $, the mean number of photons at time $t$ is given by
\begin{equation}
\begin{array}{l}
{\left\langle {{\hat n}(t)} \right\rangle} = 0.25\left[ {f_1^2(t)\left( {1 + \kappa \int_0^t {dt'g_1^2(t')} } \right)} \right.\\
\qquad\qquad\qquad \left. { + f_2^2(t)\left( {1 + \kappa \int_0^t {dt'g_2^2(t')} } \right) - 2} \right] .
\end{array}
\label{Eq9}
\end{equation}
In order to get a clear insight into Eq. (\ref{Eq9}), we rewrite it as follows
\begin{equation}
\begin{array}{l}
{\left\langle {{\hat n}(t)} \right\rangle} = 0.25\left[ {\left( {{e^{ - {\gamma _ + }t - z \sin(\Omega t)}} + {e^{ - {\gamma _ - }t + z \sin(\Omega t)}} - 2} \right)} \right.\\
\qquad \qquad \qquad \qquad \left. { + \kappa \left( {{I_ + }(t) + {I_ - }(t)} \right)} \right] ,
\end{array}
\label{Eq10}
\end{equation}
where $z=2\tilde \eta $, $ {\gamma _ \pm } = \kappa \pm 2\varepsilon $, and
${I_\pm }(t) ={{\tilde I}_ \pm }(t){e^{ - {\gamma _ \pm }t \mp 2z\sin(\Omega t)}}$ with
\begin{equation}
\begin{array}{l}
{{\tilde I}_ \pm }(t) = \int_0^t \left\{ \begin{array}{l}
g_2^2(t')\\
g_1^2(t')
\end{array} \right\}dt' =\\
\qquad\qquad \sum\limits_{m =  - \infty }^\infty  {\frac{{{{( - 1)}^m}{J_m}(z)}}{{{\gamma _ \pm } + im\Omega }}} \left( {{e^{({\gamma _ \pm } + im\Omega )t}} - 1} \right) ,
\end{array}
\label{Eq11}
\end{equation}
in which ${{J_{m}}}$ is the Bessel function of the first kind of integral order $m$. By using the properties of the Bessel function\cite{arfcen} we obtain \\
\begin{widetext}
\begin{equation}
\begin{array}{l}
{I_ \pm }(t) = {e^{ \mp z\sin(\Omega t)}}\left\{ {{w_ \pm }{J_0}(iz)\left( {{e^{ - {\gamma _ \pm }t}} - 1} \right) + \sum\limits_{m = 0}^\infty  {\frac{{2{J_{2m}}(iz)}}{{\gamma _ \pm ^2 + 4{m^2}{\Omega ^2}}}\left[ {2m\Omega \sin(2m\Omega t) + {\gamma _ \pm }\left( {\cos(2m\Omega t) - {e^{ - {\gamma _ \pm }t}}} \right)} \right]} } \right.\\
\qquad\qquad \qquad \qquad \left. { \mp \sum\limits_{m = 0}^\infty  {\frac{{2i{J_{2m + 1}}(iz)}}{{\gamma _ \pm ^2 + {{(2m + 1)}^2}{\Omega ^2}}}\left[ {{\gamma _ \pm }\sin((2m + 1)\Omega t) + (2m + 1)\Omega \left( {{e^{ - {\gamma _ \pm }t}} - \cos((2m +1 )\Omega t)} \right)} \right]} } \right\} ,
\end{array}
\end{equation}
\label{Eq12}
\end{widetext}
where ${w_ \pm } = {1}/{{{\gamma _ \pm }}}$. In this manner, the mean number of photons can be separated into three contributions as 
\begin{equation}
\left\langle {\hat n(t)} \right\rangle  = {n_{OPA}}+ {n_\eta } + {n_{interference}} = {n_{OPA}} + {n_{Casimir}},
\label{Eq13}
\end{equation}
where
\begin{equation}
{n _{OPA}} = \frac{{2{\varepsilon ^2}}}{{{\gamma _ + }{\gamma _ - }}} + \frac{\varepsilon }{{2{\gamma _ + }}}{e^{ - {\gamma _ + }t}} - \frac{\varepsilon }{{2{\gamma _ - }}}{e^{ - {\gamma _ - }t}},
\label{nOPA}
\end{equation}
\begin{equation}
{n_\eta } = \frac{1}{4}\kappa \left( {{J_0}(iz) - 1} \right)\left[ {\frac{1}{{{\gamma _ + }}} + \frac{1}{{{\gamma _ - }}} - \left( {\frac{{{e^{ - {\gamma _ + }t}}}}{{{\gamma _ + }}} + \frac{{{e^{ - {\gamma _ - }t}}}}{{{\gamma _ - }}}} \right)} \right],
\end{equation}
\label{Eq15} 
\begin{widetext}
\begin{equation}
\begin{array}{l}
4{n_{interference}}= \left( {{e^{z\sin(\Omega t)}} - 1} \right)\left( {{e^{ - {\gamma _ - }t}} + \frac{{\kappa {J_0}(iz)}}{{{\gamma _ - }}}\left( {1 - {e^{ - {\gamma _ - }t}}} \right)} \right) + \left( {{e^{ - z\sin(\Omega t)}} - 1} \right)\left( {{e^{ - {\gamma _ + }t}} + \frac{{\kappa {J_0}(iz)}}{{{\gamma _ + }}}\left( {1 - {e^{ - {\gamma _ + }t}}} \right)} \right)\\
\qquad \qquad \qquad + \kappa \sum\limits_{ + , - } {\left( {{e^{ \mp z\sin(\Omega t)}}\left\{ {\sum\limits_{m = 1}^\infty  {\frac{{2{J_{2m}}(iz)}}{{\gamma _ \pm ^2 + 4{m^2}{\Omega ^2}}}\left[ {2m\Omega \sin(2m\Omega t) + {\gamma _ \pm }\left( {\cos(2m\Omega t) - {e^{ - {\gamma _ \pm }t}}} \right)} \right]} } \right.} \right.} \\
\qquad \qquad \qquad \left. {\left. { \mp \sum\limits_{m = 1}^\infty  {\frac{{2i{J_{2m - 1}}(iz)}}{{\gamma _ \pm ^2 + {{(2m - 1)}^2}{\Omega ^2}}}\left[ {{\gamma _ \pm }\sin((2m - 1)\Omega t) + (2m - 1)\Omega \left( {{e^{ - {\gamma _ \pm }t}} - \cos((2m - 1)\Omega t)} \right)} \right]} } \right\}} \right) .
\end{array}
\label{Eq16}
\end{equation}
\end{widetext}
As can be seen, ${n_{interference}}$ is an oscillatory function of time and it can be shown that in the sum over $m$ in Eq. (\ref{Eq16}) the term corresponding to $\textit{m}=1$ has the dominant contribution. We remind that since OPA is operating below threshold, the parameters $\varepsilon$ and $\eta $ are constrained by the inequality $\kappa  > 2(\varepsilon  + \eta )$. In the absence of the time modulation, i.e., $\eta=0$, both the terms $n_\eta$ and ${n_{interference}}$ vanish and the mean number of generated photon reduces to $n_{OPA}$ given by Eq.(\ref{nOPA}). In the steady state, $n_{OPA}$ reaches $(2{\varepsilon ^2})/({\kappa ^2} - 4{\varepsilon ^2})$. Figure \ref{Fig2}(a) illustrates $n_{OPA}$ versus the dimensionless time $\kappa t$ for different values of $\varepsilon/\kappa$. As is seen, with increasing $\varepsilon/\kappa$ towards its maximum value($<0.5$) the rate of photon generation increases. In Figs. \ref{Fig2}(b)-(d) we have plotted $\ n_{Casimir}={n_\eta } + {n_{interference}} $ as a function of the dimensionless time $\Omega t$ for different values of $\kappa/\Omega$ (with a fixed value of $\kappa$). As can be seen, $\ n_{Casimir}$ exhibits an oscillatory behaviour in the course of time. The oscillations can be attributed to the CRTs which manifest themselves as an interference between the constant and time varying parts of the external pumping $\varepsilon_p$. For a given value of $\varepsilon/\eta$, decreasing the modulation frequency $\Omega$(or equivalently, increasing $\kappa/\Omega$ for a fixed value of $\kappa$) increases the amplitude of oscillations. In addition, it is seen that for a fixed value of  $\kappa/\Omega$  the mean number of generated Casimir photons increases with increasing $\varepsilon/\eta$. Figures \ref{Fig2}(b)-(d) also show that, depending on the values of $\kappa/\Omega$ and $\varepsilon/\eta$, the mean number of Casimir photons takes negative values within some time intervals, which means that $\left\langle {\hat n(t)} \right\rangle  < {n_{OPA}}$, although we have always $\left\langle {\hat n(t)} \right\rangle \ge 0$. This effect can be described as a coherent annihilation of pair photons due to the time modulation of the external pumping. With increasing the value of $\Omega$, the mean number of annihilated photons decreases and for a fixed value of $\Omega$ this annihilation occurs for large values of $\varepsilon/\eta$. The effect of the photon creation and annihilation can be attributed to the exchange of the energy between the external modulated pump and the quantum vacuum fluctuations. If the energy is transferred to the quantum vacuum fluctuations, photon pairs are created. Conversely, if the energy is gained from the quantum vacuum, photon pairs are annihilated. Indeed, the term $\cos(\Omega t)$ corresponds to the energy exchange such that the external pump varies between $\varepsilon-\eta$ and $\varepsilon+\eta$. It is interesting to note that this annihilation effect has also been reported very recently \cite{Microscopic toy model} in a microscopic toy model for cavity DCE consisting of a non-stationary dielectric slab inside a fixed single-mode cavity. The slab, which is modelled as a set of $N$ non-interacting atoms, oscillates according to an external law, while its dielectric properties are modulated externally via electric or magnetic fields. It has been shown that for $N=1$, the modulation of atomic parameters can lead to coherent annihilation of three photons accompanied by the creation of one atomic excitation. The authors has called this effect ``Anti-DCE".

An advantage of the system discussed so far, in comparison to the proposed system in \cite{44}, is the controllability of the mean number of generated Casimir photons through the external parameters $\Omega, \varepsilon$, and $\eta$.  

\section{\label{sec4}QUANTUM STATISTICAL PROPERTIES AND POWER SPECTRUM OF GENERATED PHOTONS}
In this section, we study the quantum statistical properties of the generated photons, including quadrature squeezing and counting statistics. We also calculate the power spectrum. In particular, we will show how one can manipulate these properties through controlling the parameters of the external modulated pumping.
\subsection{Quadrature Squeezing}

\begin{figure*}
\begin{center}
\includegraphics[scale=1]{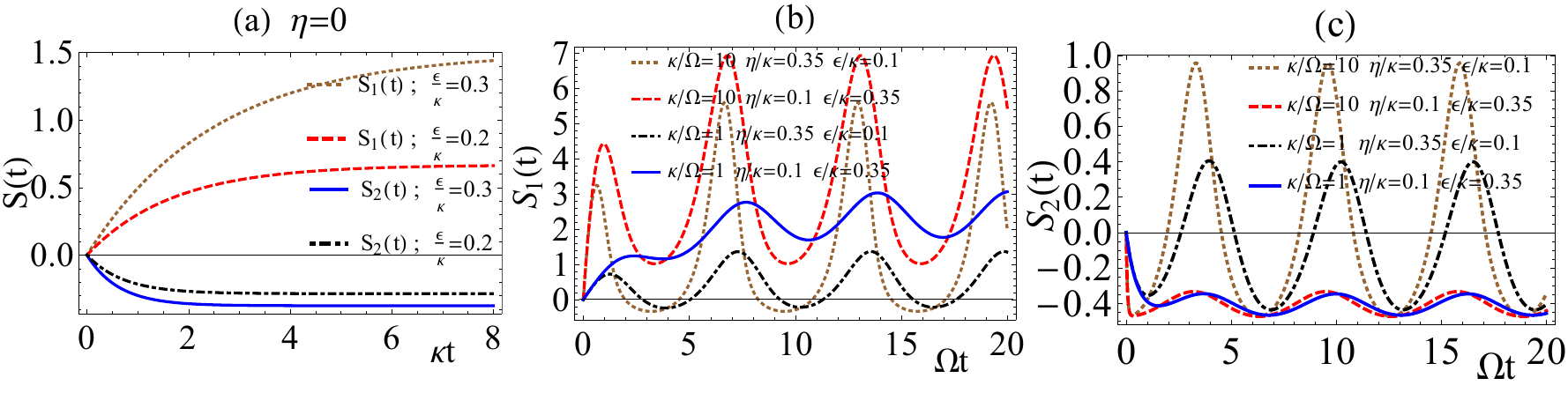}
\end{center}
\caption{(Color online) (a) The time evolution of the squeezing parameters $S_1(t)$ and $S_2(t)$ versus the dimensionless time $\kappa t$ in the case of $\eta=0$ (constant pumping) and for two values of $\varepsilon/\kappa$. (b) The time evolution of the squeezing parameter $S_1(t)$ versus the dimensionless time $\Omega t$ for the case of $\eta  \ne 0$ (time-modulated pumping) and for different values of $\kappa/\Omega$, $\varepsilon/\kappa$, and $\eta/\kappa$. (c) The time evolution of the squeezing parameter $S_2(t)$ versus the dimensionless time $\Omega t$ for the case of $\eta  \ne 0$ (time-modulated pumping) and for different values of $\kappa/\Omega$, $\varepsilon/\kappa$, and $\eta/\kappa$. Here, we have set $\kappa / 2\pi=10^4$ Hz.
}
\label{Fig3}
\end{figure*}
The squeezing property of the radiation field can be analysed by calculating the variances of the quadrature operators. Using Eq.(\ref{Eq7}) and properties of the noise operator ${{\hat F}_c}(t)$, given by Eqs.(\ref{Eq3}) and (\ref{Eq4}), the variances of the quadrature operators $\hat {a_1}$ and $\hat {a_2}$ are, respectively, obtained as
\begin{equation}
\begin{array}{l}
{(\Delta \hat a_1(t))}^2 = \frac{1}{4}f_1^2(t)\left( {1 + \kappa \int_0^t {dt'g_1^2(t')} } \right),\\
{(\Delta \hat a_2(t))}^2 = \frac{1}{4}f_2^2(t)\left( {1 + \kappa \int_0^t {dt'g_2^2(t')} } \right),
\end{array}
\label{Eq17}
\end{equation}
where ${{\hat a}_1} = \hat X/2$ and ${{\hat a}_2} = i\hat Y/2$ are quadrature operators and obey the commutation relation $\left[ {{{\hat a}_1},{{\hat a}_2}} \right] = i/2$. A state of the radiation field is said to be squeezed whenever ${(\Delta \hat a_i(t))}^2<0.25$, ($i=1$, 2). The degree of squeezing can be measured by the squeezing parameter $S_i$ ($i=1$, 2), defined by ${S_i}(t) = 4{(\Delta {{\hat a}_i}(t))^2} - 1$. Then, the condition for squeezing in the quadrature component can be simply written as $S_i(t)<0$.
Using Eq.(\ref{Eq11}) the different contributions in Eq.(\ref{Eq17}) can be separated in a similar manner as the case of photon number in the previous section. In Fig. \ref{Fig3}(a) we have plotted the squeezing parameters $S_1$ and $S_2$ versus the dimensionless time $\kappa t$ for the case of the constant pumping($\eta=0$) and for two values of $\varepsilon/\kappa$. As is seen, the quadrature component $\hat a_2$ exhibits squeezing at all times, while the component $\hat a_1$ is never squeezed. Furthermore, as time goes on, the squeezing parameters $S_1$ and $S_2$ tend to asymptomatic values $2\varepsilon /(\kappa  - 2\varepsilon )$ and $- 2\varepsilon /(\kappa  + 2\varepsilon )$, respectively. Numerical results corresponding to the temporal evolution of the squeezing parameters $S_1$ and $S_2$ in the case of time-modulated pumping($\eta  \ne 0$), for different values of $\kappa/\Omega$, $\varepsilon/\kappa$ and $\eta/\kappa$ ($\kappa$ being constant) are shown in Figs. \ref{Fig3}(b) and \ref{Fig3}(c), respectively. The effect of CRTs leads to the oscillatory behaviour of quadrature squeezing in the course of time such that the dominant frequency of these oscillations depends on $\Omega $. Unlike the constant pumping(Fig. \ref{Fig3}(a)), in this case both quadratures ${\hat a}_{1}$ and ${\hat a}_{2}$ can be squeezed in some time intervals (see Figs. \ref{Fig3}(b), (c)). It should be reminded that this effect is due to the CRTs. As shown in Figs.\ref{Fig3}(b) and(c), the quadrature squeezing can be controlled by the modulation frequency, $\Omega$, as an external control parameter. 
For fixed values of $\kappa$, $\eta$, and $\varepsilon$ when $\Omega$ decreases ($\kappa/\Omega$ increases) the maximum degree of squeezing and the amplitude of oscillations increase. Furthermore, for a fixed value of $\kappa/\Omega$, when $\varepsilon/\eta$ increases, the degree of squeezing of ${\hat a}_2$ increases while the squeezing of the component ${\hat a}_1$ is suppressed. Our calculations show that if $\kappa/\Omega$ be very smaller than unity, the temporal behaviours of the quadrature variances are similar to those obtained for the OPA but with very small oscillations. By controlling the modulation frequency $\Omega$ the maximum quadrature squeezing is about 0.13, i.e., $50\%$ noise reduction is achievable. We remind that we have plotted the squeezing parameters for the total generated photons. However, it is easy to show that the quadrature variances for generated Casimir photons are
\begin{equation}
\begin{array}{l}
{(\Delta {{\hat a}_{i,Casimir}})^2} = {(\Delta {{\hat a}_i})^2} - {(\Delta {{\hat a}_{i,\varepsilon }})^2}\\
\qquad\qquad\qquad = {(\Delta {{\hat a}_{i,\eta }})^2} + {(\Delta {{\hat a}_{i, {interference}}})^2},
\end{array} 
\label{casimir-squeezing}
\end{equation}
where ${(\Delta {{\hat a}_{i,\varepsilon }})^2}$, ${(\Delta {{\hat a}_{i,\eta }})^2}$, and ${(\Delta {{\hat a}_{i,{interference}}})^2}$ are, respectively, the contribution of constant pumping, the time modulated term and the interference term which is due to the CRTs. Numerical results reveal (not shown here) that the behaviour of ${(\Delta {{\hat a}_{i,Casimir}})^2}$ is almost similar to that of ${(\Delta {{\hat a}_i})^2}$, that is the difference is quantitative rather than qualitative.

\subsection{PHOTON COUNTING STATISTICS: Autocorrelation function and the Mandel parameter}
\begin{figure}
\begin{center}
\includegraphics[scale=1]{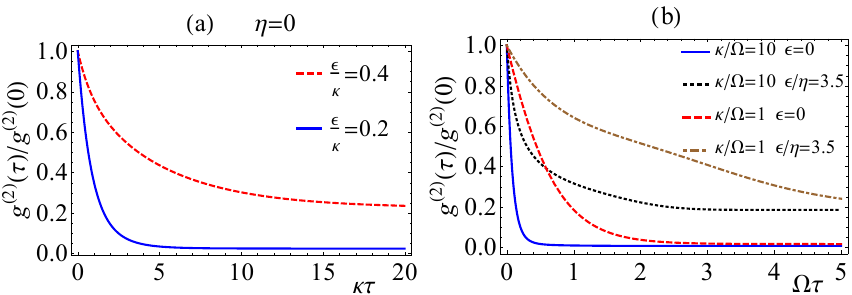}
\end{center}
\caption{(Color online) (a) Normalized autocorrelation function ,${g^{(2)}}(\tau )/{g^{(2)}}(0)$ , versus the scaled delay time $\kappa \tau$ in the absence of time modulation and for different values of ${\varepsilon }/{\kappa }$. (b) Normalized autocorrelation function , ${g^{(2)}}(\tau )/{g^{(2)}}(0)$, versus the scaled delay time $\Omega \tau$ in the presence of time-modulated pumping and for different values of $\kappa/\Omega$ and $\varepsilon/\eta$. Here, we have set $\kappa/2 \pi = 10^4$ Hz.
} 
\label{Fig4}
\end{figure}

\begin{figure*}
\begin{center}
\includegraphics[scale=1]{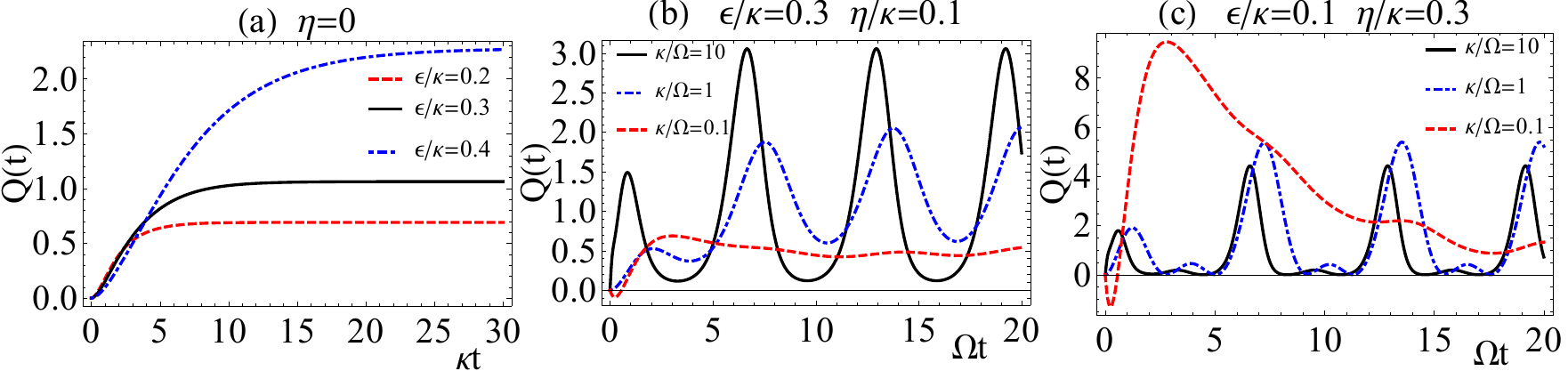}
\end{center}
\caption{(Color online) (a) Time evolution of the Mandel parameter, $Q(t)$, as a function of the dimensionless time $\kappa t$, in the absence of time modulation for different values of $\varepsilon/\kappa$. (b) Time evolution of the Mandel parameter, $Q(t)$, as a function of the dimensionless time $\Omega t$, in the presence of time modulation for $\varepsilon/\kappa=0.3$ and $\eta/\kappa=0.1$ and different values of $\kappa/\Omega$. (c) Time evolution of the Mandel parameter, $Q(t)$, as a function of the dimensionless time $\Omega t$, in the presence of time modulation for $\varepsilon/\kappa=0.1$ and $\eta/\kappa=0.3$ and for different values of $\kappa/\Omega$. Here, we have set $\kappa / 2\pi=10^4$ Hz.
}
\label{Fig5}
\end{figure*}

The autocorrelation function is proportional to the probability of detecting one photon at time $t + \tau $ given that another photon was detected at earlier time $t$. It is defined by
\begin{equation}
g^{(2)}{(\tau )} = \frac{{{{\left\langle {\hat a^\dag{(t)} \hat a^\dag{(t + \tau )} {{\hat a}{(t + \tau )}}{{\hat a}{(t)}}} \right\rangle }}}}{{\left\langle {\hat a^\dag{(t)} {{\hat a}{(t)}}} \right\rangle ^2}}.
\label{Eq18}
\end{equation}
When the radiation field satisfies the inequality ${g^{(2)}}(\tau ) < {g^{(2)}}(0)$, the photons tend to distribute themselves preferentially in bunches rather than at random(photon bunching). On the other hand, if ${g^{(2)}}(\tau ) > {g^{(2)}}(0)$ fewer photon pairs are detected close together than further apart (photon antibunching). Using the Gaussian properties of the noise forces and the quantum regression theorem \cite{scully, milburn} we get, after algebraic manipulations, the expression of autocorrelation function
 \begin{equation}
g^{(2)} {(\tau )}= 1 + \frac{{{{\left| {{{\left\langle {\hat a^\dag{(t)} \hat a^\dag{(t + \tau )} } \right\rangle }}} \right|}^2}}}{{\left\langle {\hat a^\dag{(t)} {{\hat a}{(t)}}} \right\rangle^2}} + \frac{{{{\left| {{{\left\langle {\hat a{(t)}^\dag {{\hat a}{(t + \tau )}}} \right\rangle }}} \right|}^2}}}{{\left\langle {\hat a^\dag{(t)}{{\hat a}{(t)}}} \right\rangle^2}}.
 \label{Eq19}
 \end{equation}
In order to determine two correlation functions appearing in the right hand side of Eq. (\ref{Eq19}) one can use Eq. (\ref{Eq7}) together with the correlation functions for the noise operators ${{\hat F}_C}$, given by Eq. (\ref{Eq3}) and (\ref{Eq4}). The results are as follows 
\begin{widetext}
\begin{equation}
\begin{array}{l}
\left\langle {{{\hat a}^\dag }(t)\hat a(t + \tau )} \right\rangle  = 0.25\left[ {{f_1}(t){f_1}(t + \tau )\left( {1 + \kappa \int_0^t {dt'g_1^2(t')} } \right) + {f_2}(t){f_2}(t + \tau )\left( {1 + \kappa \int_0^t {dt'g_2^2(t')} } \right)} \right.\\
\qquad\qquad\qquad\qquad\qquad\qquad\qquad \left. { - 2{e^{ - 0.5\kappa \left| \tau  \right|}}Cosh\left( {\varepsilon \tau  + 2\frac{\eta }{\Omega }Sin\frac{{\Omega \tau }}{2}Cos\frac{{\Omega (2t + \tau )}}{2}} \right)} \right]
\end{array}
 \label{Eq20}
 \end{equation}
 \begin{equation}
\begin{array}{l}
\left\langle {\hat a^\dag {(t)}\hat a^\dag{(t + \tau )} } \right\rangle  = 0.25\left[ {{f_1}(t){f_1}(t + \tau )\left( {1 + \kappa \int_0^{t} {dt'g_1^2(t')} } \right) + {f_2}(t){f_2}(t + \tau )\left( {\kappa \int_0^{t} {dt'g_2^2(t') - 1} } \right)} \right.\\
\qquad    \qquad   \qquad   \qquad    \qquad \qquad \qquad \left. { - 2{e^{ - 0.5\kappa \left| \tau  \right|}}\sinh\left( {\varepsilon \tau  + 2\frac{\eta }{\Omega }\sin(\frac{{\Omega \tau }}{2})\cos(\frac{{\Omega (2t + \tau )}}{2})} \right)} \right] ,
\end{array}
 \label{Eq21}
 \end{equation}
\end{widetext}
Considering these two expressions for times much longer than the cavity relaxation time ($t \gg {\kappa ^{ - 1}}$) we have plotted  ${g^{(2)}}(\tau )/{g^{(2)}}(0)$ in Fig. \ref{Fig4}(a) for the case of constant pumping ($\eta = 0)$ as a function of $\kappa \tau$, for two values of $\varepsilon/\kappa$ ($\kappa=$ constant). As is seen, in this case, the generated photons exhibit bunching phenomenon. Also, decreasing $\varepsilon$ leads to an inhibition of photon bunching. In Fig. \ref{Fig4}(b), we have plotted ${g^{(2)}}(\tau )/{g^{(2)}}(0)$ for the case of time-modulated pumping ($\eta \ne 0$) as a function of $\Omega \tau$ for different values $\kappa/\Omega$ and $\varepsilon/\eta$. As is seen, for a fixed value of $\kappa/\Omega$ when $\varepsilon/\eta$ increases photon bunching decreases and ${g^{(2)}}(\tau )/{g^{(2)}}(0)$ approaches 0.3. Furthermore, with increasing $\kappa/\Omega$ (decreasing the modulation frequency $\Omega$) when $\kappa$ is constant, photon bunching increases and ${g^{(2)}}(\tau )/{g^{(2)}}(0)$ approaches 0. The fast decay of the autocorrelation function is due to the CRTs. Numerical analysis shows that  If ${\kappa }/{\Omega }<1$, the behaviour of the autocorrelation function is similar to that obtained for the OPA, but it has oscillatory behaviour at long delay times. Increasing $\varepsilon/\eta$ leads to the decreasing of the amplitude of these oscillations. The manifestation of photon bunching is attributed to the possibility of emission of two photons simultaneously by DCE that is an OPA like process (it can be described by the Hamiltonian of Eq.(\ref{Eq1})). Up to now, we have found that for a fixed value of $\kappa$ with the decreasing value of $\Omega$, the mean number of generated Casimir photons, the maximum degree of squeezing and photon bunching increase. To determine the photon statistics, one can calculate the Mandel parameter which is defined as $Q = ((\Delta {{n})^2} - \left\langle {\hat n} \right\rangle )/\left\langle {\hat n} \right\rangle  = (\left\langle {{{\hat a}^{\dag 2}}{{\hat a}^2}} \right\rangle  - {\left\langle {\hat n} \right\rangle ^2})/\left\langle {\hat n} \right\rangle$ where
\begin{equation}
\left\langle {{{\hat a}^{\dag 2}}{{\hat a}^2}} \right\rangle  = \sum\limits_{i = 1}^8 {{A_i}(t) {F_i}(t)} .
\label{Eq.ad2a2}
\end{equation}
the functions $F_i(t)$ and $A_i(t)$ are as follows
\begin{equation}
\begin{array}{l}
{A_1}(t) = \frac{1}{{16}}\left[ {10{e^{ - 2\kappa t}} + 3\left( {{f_1}^4(t) + {f_2}^4(t)} \right)} \right.\\
\qquad\qquad\qquad\left. { - 8{e^{ - \kappa t}}\left( {{f_1}^2(t) + {f_2}^2(t)} \right)} \right],\\
{A_2}(t) = {A_5}(t) = {A_6}(t) = {A_7}(t) = \frac{1}{4}\left( {{f_1}^2(t) - {f_2}^2(t)} \right),\\
{A_3}(t) = \frac{1}{4}\left( {{f_1}^2(t) - {f_2}^2(t) - 2{e^{ - \kappa t}}} \right),\\
{A_4}(t) = \frac{1}{4}\left( {{f_1}^2(t) + {f_2}^2(t) - 2{e^{ - \kappa t}}} \right),\\
{A_8}(t) = 1,\\
\end{array}
\label{Eq.Ai}
\end{equation}
\begin{equation}
\begin{array}{l}
{F_1}(t) = 1,\\
{F_2}(t) = {F_7}(t) = \frac{\kappa }{4}\left( {f_1^2(t)\int_0^t {g_1^2(x)dx - f_2^2(t)\int_0^t {g_2^2(x)dx} } } \right),\\
{F_3}(t) = {F_4}(t) = {F_5}(t) = {F_6}(t) = \frac{1}{4}\left[ {\kappa f_1^2(t)\int_0^t {g_1^2(x)} } \right.\\
\qquad\qquad\qquad\qquad\left. { + \kappa f_2^2(t)\int_0^t {g_{12}^2(x)} dx - 2(1 - {e^{ - \kappa t}})} \right],\\
{F_8}(t) = {F_7}{F_2} + {F_6}{F_3} + {F_5}{F_4} .
\end{array}
\label{Eq.Fi}
\end{equation}
For $Q>0$ ($Q<0$), the statistics is super-Poissonian (sub-Poissonian); $Q=0$ stands for Poissonian statistics.
Figure \ref{Fig5}(a) shows the temporal evolution of the Mandel parameter in the absence of time modulation ($\eta=0$) versus the scaled time $\kappa t$ for different values of $\varepsilon/\kappa$. As is seen, as time goes on, the Mandel parameter increases and it is finally stabilized at an asymptotic value. With the increasing value of $\varepsilon$ the super-Poissonian statistics is enhanced. Furthermore, the rates with which the reaching the  asymptotic values occur are inversely proportional to $\varepsilon$; the smaller $\varepsilon$ is, the more rapidly the Mandel parameter tends to the asymptotic value.
Figures \ref{Fig5}(b) and (c) show the temporal evolution of the Mandel parameter for the generated Casimir photons versus the scaled time $\Omega t$ for different values of $\kappa/\Omega$, $\varepsilon/\kappa$ and $\eta/\kappa$. As is seen, for $\kappa/\Omega >1$, $Q(t)>0$ for all times, which means that the generated Casimir radiation obeys super-Poissonian statistics. The Mandel parameter exhibits an oscillatory behaviour due to the CRTs such that the frequency of oscillations depends on $\Omega$. 
Furthermore, we observe that the super-Poissonian behaviour can be controlled by $\Omega$ and $\varepsilon/\eta$ in such a way that if $\kappa$ and $\Omega$ are constant, by increasing $\varepsilon/\eta$ one not only decreases the amplitude of the oscillations of the Mandel parameter, but also decreases its maximum value. Furthermore for a fixed value of $\varepsilon/\eta<1$, increasing of $\kappa/\Omega$($\kappa=$constant) leads to decreasing of the amplitude of the oscillations in the Mandel parameter and its maximum value(Fig.\ref{Fig5}(c)). On the other hand, for $\varepsilon/\eta>1$, increasing of $\kappa/\Omega$($\kappa=$constant) leads to increasing of the amplitude of the oscillations in the Mandel parameter and its maximum value(Fig.\ref{Fig5}(b)). For $\kappa/\Omega \ll 1$, the Mandel parameter has not a clear oscillatory behaviour. In this case, as is seen, at the very initial stages of evolution ($\kappa t \ll 1$), $Q(t)<0$, which means that the generated Casimir radiation obeys sub-Poissonian statistics but then it increases towards positive values (super-Poissonian).

\subsection{ Power Spectrum}

\begin{figure*}
\begin{center}
\includegraphics[scale=1]{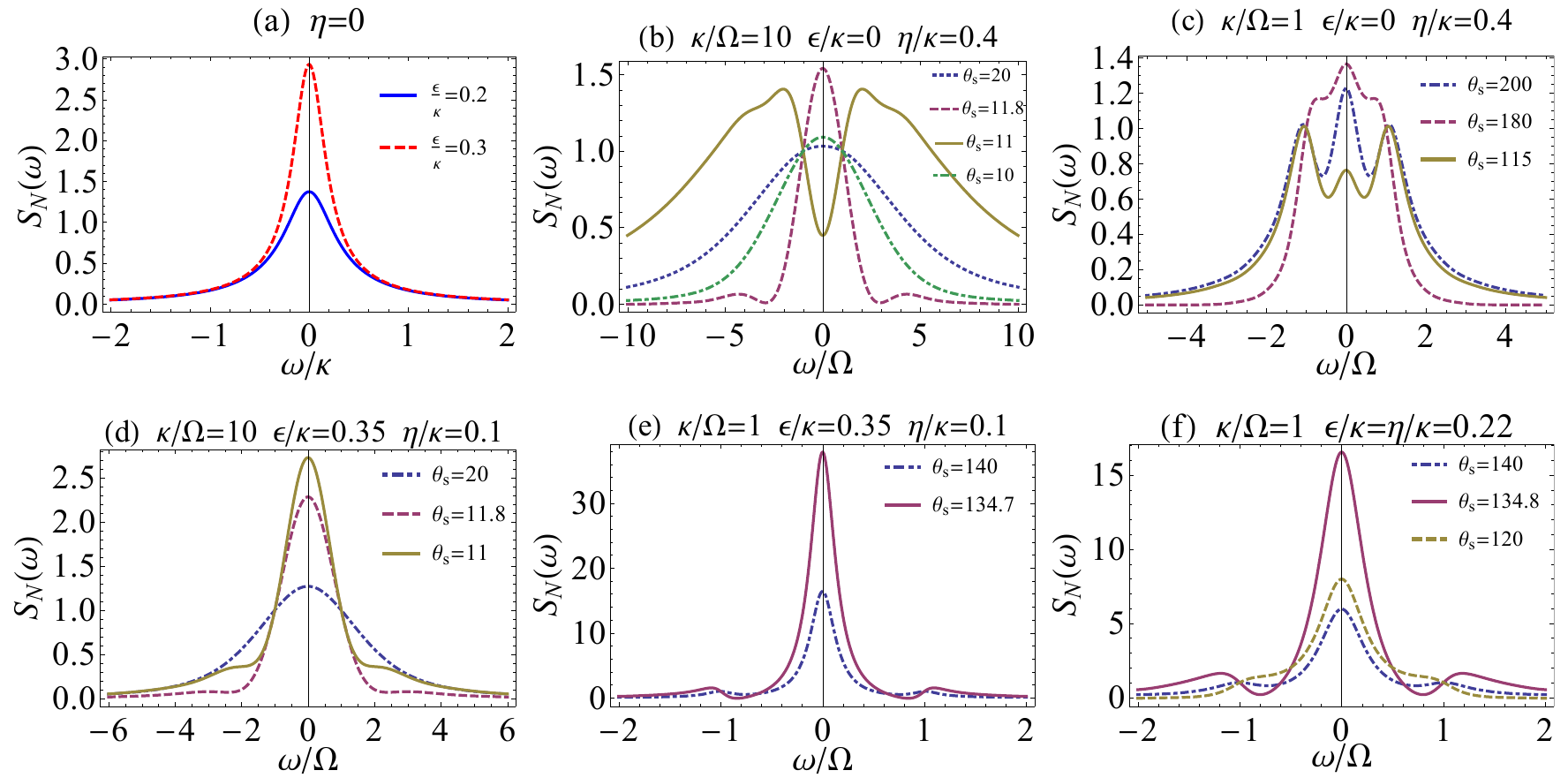}
\end{center}
\caption{(Color online) (a) Normalized power spectrum, ${S_N}(\omega ) = S(\omega )/S(\varepsilon )$, in the absence of time modulation, i.e., $\eta=0$, as a function of $\omega /\kappa $ for different values of $\varepsilon/\kappa$. (b)-(f) Normalized power spectrum, ${S_N}(\omega ) = S(\omega )/S(\Omega )$ versus $\omega /\Omega $ for different values of $\varepsilon$, $\eta$, $\Omega$, and $\theta_s=\Omega t_s$. Here, we have set $\kappa/ 2\pi=10^4$ Hz.
}
\label{Fig6}
\end{figure*}

We next proceed to calculate the power spectrum of the intracavity light field. The power spectrum can be expressed in terms of  the Fourier transform of the two-time correlation function in steady state as follows
\begin{equation}
{S{(\omega )}} = \frac{1}{\pi }{\mathop{\rm Re}\nolimits} \int_0^\infty  {d\tau {e^{i\omega \tau }}} {\left\langle {\hat a^\dag{(t)} \hat a{(t + \tau )}} \right\rangle _{ss}} ,
\label{Eq22}
\end{equation}
where the two-time correlation function,${\left\langle {\hat a^\dag {(t)}\hat a{(t + \tau )}^{}} \right\rangle}$ , is given by Eq.(\ref{Eq20}). Figures \ref{Fig6}(a)-(f) illustrate the normalized power spectra for different values of $\varepsilon, \Omega, \eta$ and $\theta_s=\Omega t_s$ (scaled interaction time in the steady state) and constant value of $\kappa/2\pi=10^4 $Hz. In Fig. \ref{Fig6}(a) we have plotted the normalized power spectrum ${S_N}(\omega ) = S(\omega )/S(\varepsilon )$ versus $\Omega/\kappa$ for $\eta=0$ and for two values of $\varepsilon$. As is seen, increasing $\varepsilon$ not only causes the strength of the peak of the spectrum to increase, but also leads to narrowing of the spectrum.
In fig.\ref{Fig6}(b)-(f) we have plotted the spectrum of the generated photons inside the cavity in the presence of time modulation as a function of $\omega/\Omega$. As can be seen, for fixed value of $\kappa/\Omega$, the width of the spectrum decreases and its height increases if $\varepsilon/\eta$ increases. In general, for constant value of $\kappa$ the width of the spectrum increases if $\kappa/\Omega$ increases or the modulation frequency decreases. 
With varying the scaled time, $\theta_s$, the shape of the spectrum, i.e., its width, height and number of peaks, vary as a result of CRTs that cause the spectrum to be a function of time $t_s$. Furthermore, varying $\theta_s$ leads to the appearance of the symmetric side peaks in the spectrum with same width and height which depend on the cavity decay rate $\kappa$, constant amplitude $\varepsilon$, modulation amplitude $\eta$ and modulation frequency $\Omega$. The distance between side peaks and their locations depend on the modulation frequency. The appearance of the side bands is due to the interference terms corresponding to the CRTs. Therefore, the spectrum can be controlled by adjusting the modulation frequency and $\theta_s$. 

It may be interesting  to compare the intracavity spectrum obtained above with the spectrum of the output photonic field. In fact the input-output theory\cite{60} links the emitted output field $\hat a_{out}$ to the intracavity field $\hat a$ by a simple relation
\begin{equation}
{{\hat a}_{out}}(t) = \sqrt \kappa  \hat a(t) - {{\hat a}_{in}}(t),
\label{Eq.aout}
\end{equation}
where, $\hat a(t)$ is the intracavity light field operator and ${{\hat a}_{in}}(t) = (1/\sqrt \kappa  ){{\hat F}_c}(t)$ is the input field. After algebraic manipulations, we find 
\begin{equation}
\begin{array}{l}
\left\langle {\hat a_{out}^\dag (t){{\hat a}_{out}}(t + \tau )} \right\rangle  = \kappa \left\langle {{{\hat a}^\dag }(t)\hat a(t + \tau )} \right\rangle  - \left\langle {{{\hat a}^\dag }(t){{\hat F}_c}(t + \tau )} \right\rangle \\
\qquad\qquad - \left\langle {\hat F_c^\dag (t)\hat a(t + \tau )} \right\rangle  + \frac{1}{{\kappa  }}\left\langle {\hat F_c^\dag (t)\hat F(t + \tau )} \right\rangle .
\end{array}
\label{Eq.nout}
\end{equation} 
At zero temperature, all of the correlation functions that appear in Eq. (\ref{Eq.nout}) are zero except for the first term, and thus
\begin{equation}
\left\langle {\hat a_{out}^\dag (t){{\hat a}_{out}}(t + \tau )} \right\rangle  = \kappa \left\langle {{{\hat a}^\dag }(t)\hat a(t + \tau )} \right\rangle .
\label{Eq.nout0}
\end{equation}
This implies that the spectrum of the output field is proportional to the spectrum of the intracavity light field, i.e., ${S_{out}}(\omega ) = \kappa S(\omega )$. Therefore, the Casimir radiation can be detected from the comparison between the output field spectrum in the absence of the time modulation($\eta=0$) and the spectrum in the presence of the time modulation.

\section{\label{sec5}summary and conclusions}
In summary, we have proposed an analogue model for controlled generation of photons via the DCE in a nonlinear cavity containing a degenerate OPA which is pumped by a time-modulated field. In the system under consideration the time modulated nonlinear parametric interaction is analogous to an apparent periodic displacement of the cavity mirrors. By solving analytically the quantum Langevin equations for the intracavity field operators without using the RWA for rapidly oscillating terms of the modulation frequency, we have found that the mean number of intracavity photons consists of three contributions: constant pumping, time varying pumping, and their interference resulting from the CRTs. The results reveal that the temporal behaviour of the mean number of generated photons can be controlled via the parameters of the time-modulated pump field. We have also identified the so-called ``Anti-DCE", in which pair photons are coherently annihilated due to the time-modulated pumping. 

In addition to the generation of  Casimir photons, we have discussed analytically and numerically their quantum statistical properties including the quadrature squeezing and photon-counting statistics, We have also examined the power spectrum of the generated Casimir radiation. We have found that the generated radiation exhibits quadrature squeezing, photon bunching, and super -Poissonian statistics in the course of time, which are controllable through the parameters of the time-modulated pumping. Furthermore, the main characteristics of the power spectrum, i.e., its height, width, and number of peaks can be controlled, in particular, by adjusting the modulation frequency and the interaction time in the steady state. The results also reveal that the comparison of the output power spectrum in the presence of the time- modulated pumping with one which obtained by constant pumping provides a way for the detection of the generated Casimir photons.





\end{document}